\documentstyle[11pt,newpasp,twoside,epsf]{article}

\newcommand{\SN}{\hbox{\rm S/N}}
\newcommand{\sauron}{\texttt{SAURON}}
\newcommand{\reffig}[1]{Fig.~\ref{#1}}

\begin{document}

\title{Adaptive Spatial Binning of 2D Spectra and Images\\ Using Voronoi Tessellations}

\author{Michele Cappellari$^1$}
\affil{Leiden Observatory, Postbus 9513, 2300 RA Leiden, The Netherlands}
\author{Yannick Copin}
\affil{Institut de Physique Nucl\'eaire de Lyon, 69222 Villeurbanne, France}

\begin{abstract}
\footnotetext[1]{European Space Agency external fellow.}
We present new techniques to perform adaptive spatial binning of two-di\-men\-sio\-nal (2D) data to reach a chosen constant signal-to-noise ratio per bin. These methods are required particularly for the proper analysis of Integral Field Spectrograph (IFS) observations, but can also be used for standard photometric imagery. Various schemes are tested and compared using data obtained with the panoramic IFS \sauron.
\end{abstract}

\section{Introduction}
\label{sec:intro}

Spatially resolved astronomical observations commonly span orders of magnitude variations in the signal-to-noise (S/N) across the detector elements. For this reason data are often grouped together in the spatial direction (binned) and averaged before analyzing them. More spatial resolution is retained in the high-S/N regions compared to the low-S/N ones.
A well known example is galaxy photometry, where logarithmically spaced radial bins are often adopted. Many more pixels are used to compute the value of a galaxy profile at large radii than in the center.

Binning is essential in the case of spectroscopic observations of the stellar kinematics. In fact a minimum S/N is \emph{required} for a reliable and unbiased extraction of kinematical information from the spectra (e.g., Rix \& White 1992; van der Marel \& Franx 1993; Kuijken \& Merrifield 1993). For this reason binning is invariably used to analyze one-dimensional (1D, e.g., long slit) spectroscopic observations.
New developments with Integral Field Spectroscopy (IFS; e.g., \texttt{OASIS} on CFHT, \sauron\ on WHT, \texttt{VIMOS} on VLT, \texttt{GMOS} on Gemini) require methods to perform spatial binning of spectra in two dimensions (2D) too.

The related process of adaptively smoothing is not acceptable for our purposes since it correlates the information of different bins which complicates the quantitative interpretation of spectroscopic measurements. For the same reason smoothing is never used for the quantitative analysis of 1D spectra.

Little work has been done on the subject of adaptive 2D-binning.  Sanders \& Fabian (2001) developed an algorithm to be used with X-ray imaging data. The bins that their method produces however can contain other bins and are not compact. In the case of spectroscopic data it makes little sense to bin together spectra coming from pixels that are not close to each other and whose properties may differ considerably. Other schemes have to be developed.

\section{Formulation of the Problem}
\label{sec:generic}

We tackle here the problem of binning in the spatial direction(s). In what follows the term `pixel' refers to a given spatial element of the dataset: it can be an actual pixel of a CCD image, or a spectrum position along the slit of a long-slit spectrograph or in the field of view of an IFS.

Each pixel $i$ has an associated signal ${\cal S}_{i}$ and its corresponding noise ${\cal N}_{i}$. The pixel signal-to-noise ratio is therefore $(\SN)_{i} = {\cal S}_{i}/{\cal N}_{i}$. Our considerations do not depend on the details used to estimate these quantities, which we assume to be known beforehand for every pixel $i$. In the case of spectrography for instance, the quantity ${\cal S}_{i}$ associated to a given spectrum $S_{i}(\lambda)$ can be the average signal over a given spectral range $\Delta\lambda$: ${\cal S}_{i} = \frac{1}{\Delta\lambda}\int_{\Delta\lambda} S_{i}(\lambda)\,{\rm d}\lambda$; while the corresponding average noise can be defined as ${\cal N}^2_{i} = \frac{1}{\Delta\lambda}\int_{\Delta\lambda} N^2_{i}(\lambda)\,{\rm d}\lambda$. It is important to note that, in our case, the term `binning' will only refer here to the averaging of observations taken at different positions on the sky (i.e., different pixels), and \emph{not} along the spectral direction.

To bin in 1D one only has to make sure that the spatially binned data satisfy a minimum \SN\ requirement (or better a minimum scatter around the target \SN). In 2D (and higher dimensions) the situation is more complex and a good binning scheme has to satisfy the following requirements:
\begin{description}
\setlength{\itemsep}{0cm}
\item [Topological requirement:] the bins should at least properly tessellate the region $\Omega$ of the plane under consideration, i.e., create a partition of $\Omega$, without overlapping or holes. While this requirement is trivial to enforce in 1D, it is tricky to implement in higher dimensions, as the bin shapes must then be taken into consideration;

\item [Morphological requirement:] the bin shape has to be as `compact' (or `round') as possible, so that they are associated with a well-defined position, and correspond to the overall best spatial resolution;

\item [Uniformity requirement:] the bin \SN\ should be as uniform as possible around a target value. While a minimum \SN\ is generally required, one does not want to sacrifice spatial resolution to increase the \SN\ even further.
\end{description}

\begin{figure}[t]
    \plottwo{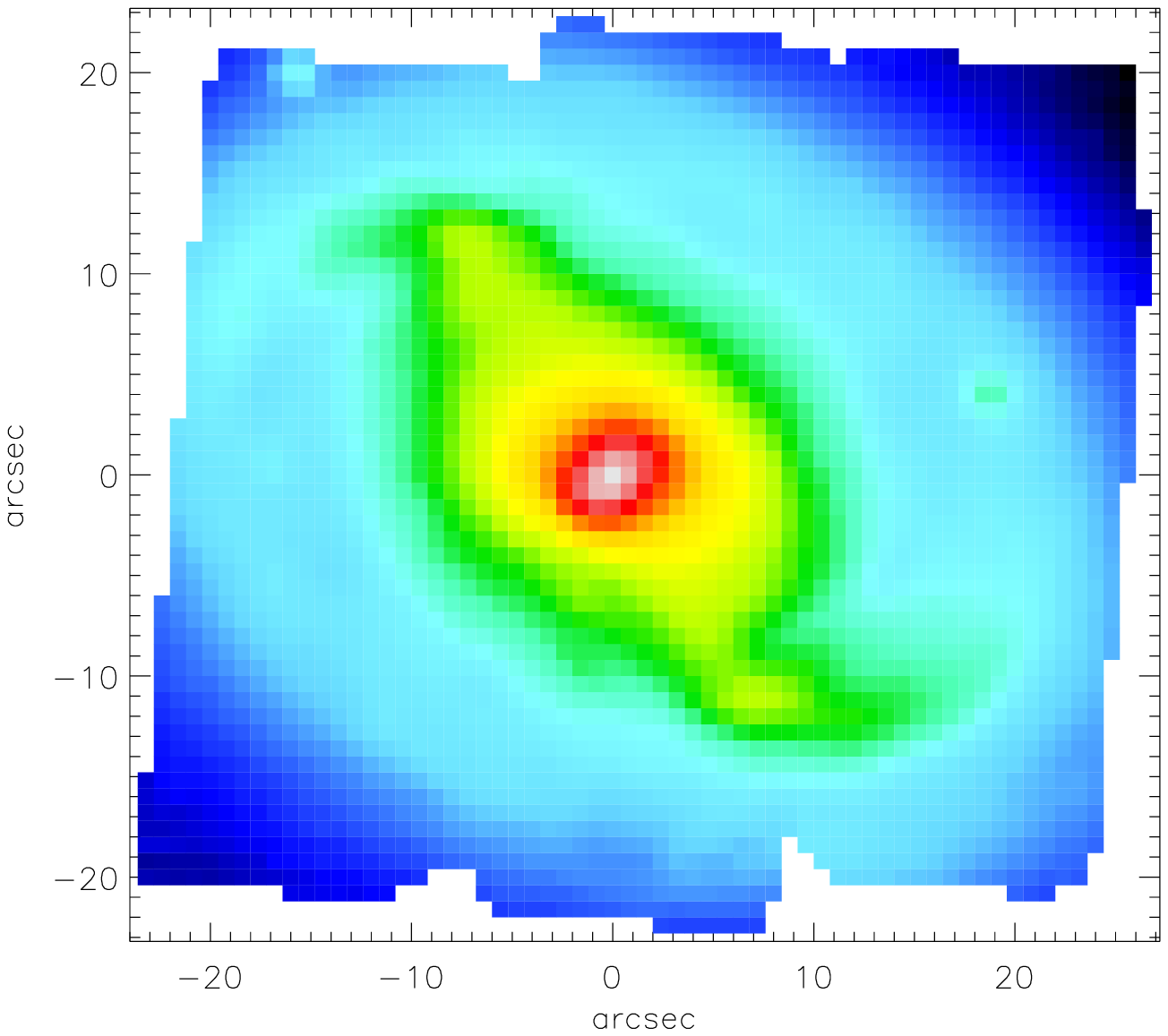}{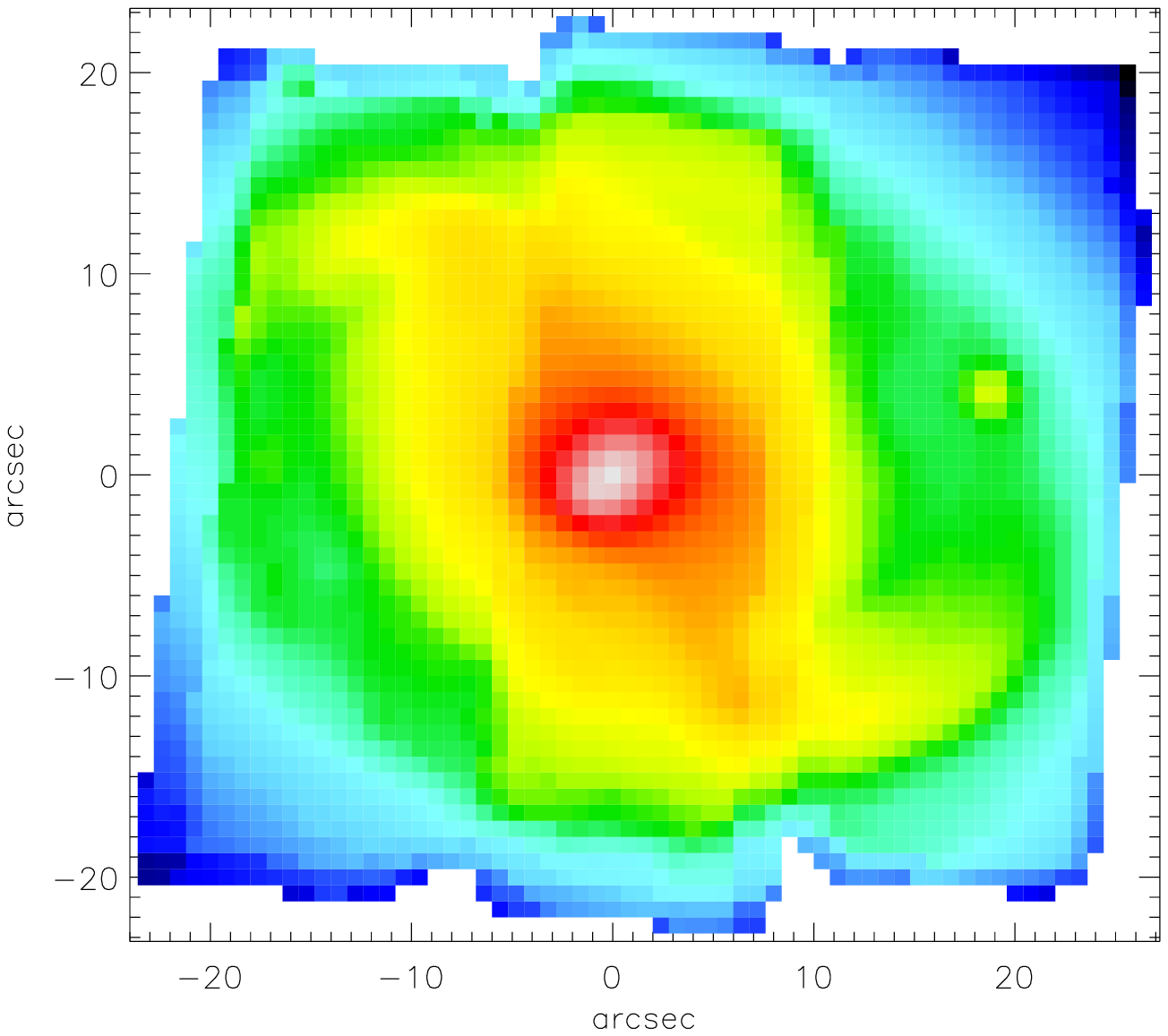}
    \caption{\sauron\ observation of the barred Sa galaxy NGC~2273, based on two $4 \times 1800$~s pointings. The field-of-view is $49'' \times 44''$, with an effective spatial sampling of $0\farcs8 \times 0\farcs8$. \emph{Left panel:} reconstructed total intensity. \emph{Right panel:} \SN-map, the \SN{} being integrated along the whole \sauron\ spectral range. Red corresponds to an average \SN$\approx$50 per resolution element. Note the galaxy spiral arms, as well as the two vertical \SN{} jumps close to the middle of the frame and the irregular boundaries due to the merging process of the different exposures. \label{fig:n2273-SNmap}}
\end{figure}

In what follows we consider different methods and we apply each one to observations of the barred Sa galaxy NGC~2273 (\reffig{fig:n2273-SNmap}) taken with the panoramic IFS \sauron\ (Bacon et al.\ 2001). These observations, based on 8 single 1800~s exposures, have been selected for having high \SN{} contrast between the inner and the outer parts, and a complex \SN{}-distribution, caused by the presence of spiral arms, \SN{}-jumps and irregular outer boundaries due to the merging of multiple exposures.

\section{Quadtree Method}
\label{sec:quadtree}

It is useful to first consider the Quadtree algorithm (Samet 1984), which we believe is close to the best `regular' image processing method available for the present application. We show that the Quadtree method cannot produce an \emph{optimal} binning and more complex schemes are required. These we discuss in Section~4.

The Quadtree method consists of a recursive partition of a region of the plane into axis-aligned squares. One square, the \emph{root}, covers the entire region. A square is divided into four \emph{child} squares, by splitting with horizontal and vertical segments through its center. The collection of squares then form a tree, with smaller squares at lower levels of the tree. The splitting of squares terminates when all squares meet some convergence criterion.

\begin{figure}[t]
    \plottwo{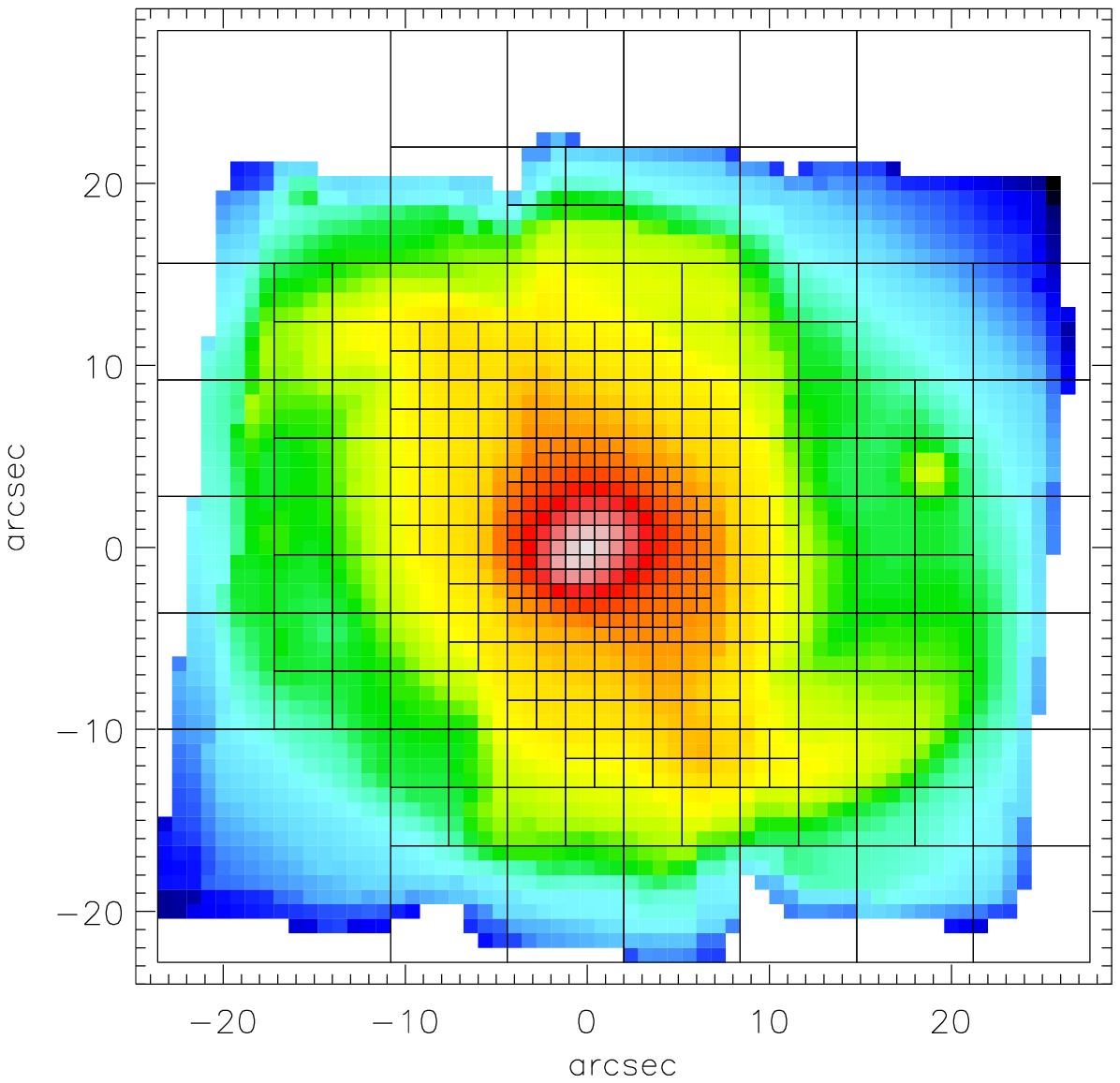}{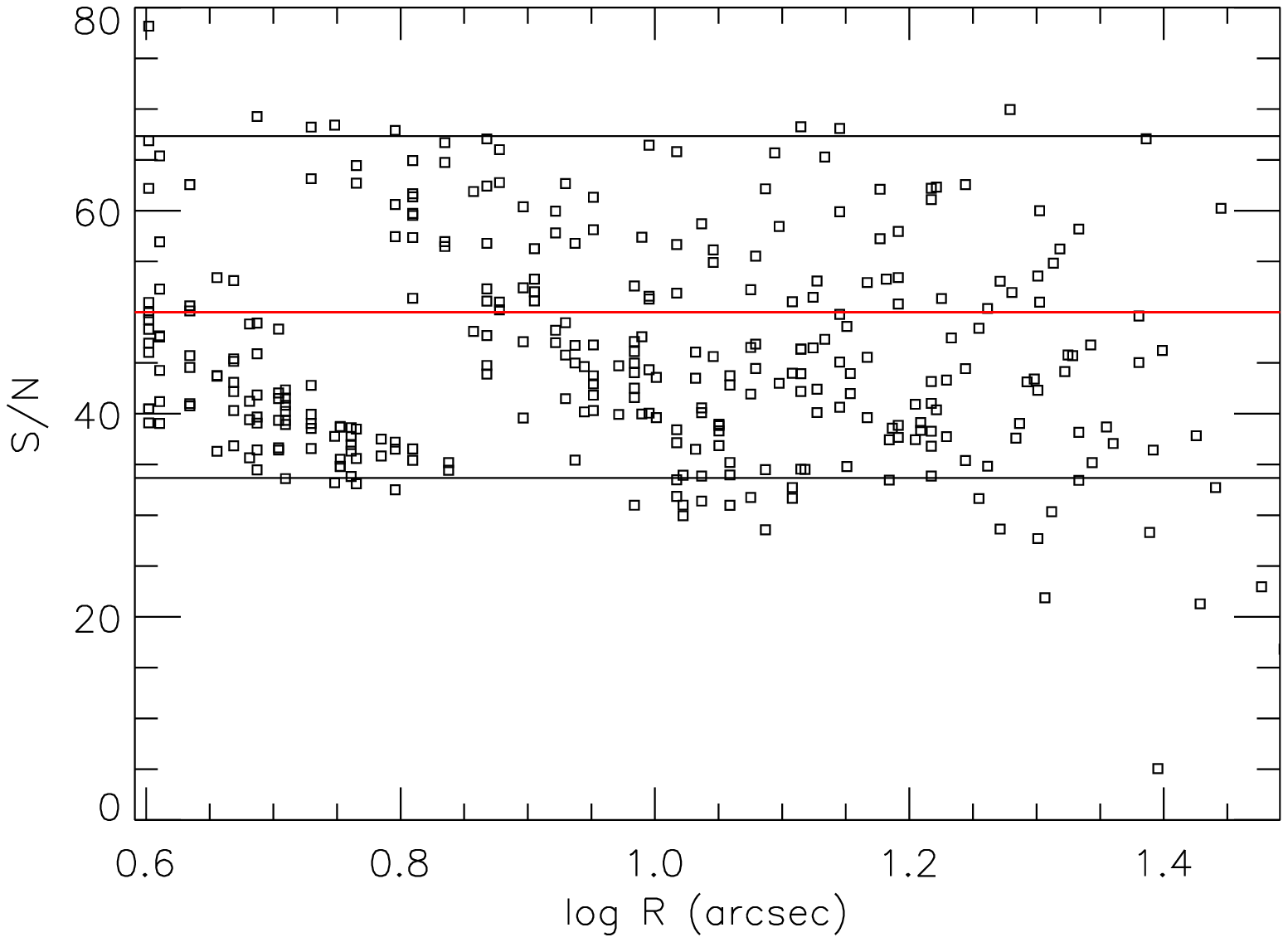}
    \caption{\emph{Left panel:} Quadtree binning of NGC~2273 for a target \SN$=$50 (red line). \emph{Right panel:} S/N variation in the Quadtree binning of NGC~2273. The open squares represent the S/N of each bin, as a function of its distance from the galaxy center. Note the factor 2$\times$ scatter in the S/N due to the factor 4$\times$ recursive splitting of the bins. The target \SN$=$50 is indicated by the red thick horizontal line.
    \label{fig:quadtree_scatter}}
\end{figure}

In \reffig{fig:quadtree_scatter} the Quadtree method was used to rebin the \SN\ map of \reffig{fig:n2273-SNmap} into squares satisfying a minimum S/N requirement. The nice feature of this binning method is that the resulting bins  are squares of various sizes (except at the border). In this way bins are easy to handle, and require little information to be described completely.

There are however two major problems with this method:
\begin{itemize}
\setlength{\itemsep}{0cm}
\item a S/N spread of a factor $\sim2$ is unavoidable due to the fact that bin area varies by construction in steps of a factor of 4;

\item unless the original image has a size which is a power of two, some bins at the border will not be square and generally will not meet the minimum S/N criterion, becoming unusable for later analysis.
\end{itemize}

\section{Voronoi Tessellation}
\label{sec:vt}

Given the inability of methods employing `regular' bins to produce optimal 2D tessellations, we now consider schemes which do not have square or rectangular bins. Accordingly we consider the \emph{Voronoi Tessellation} (VT), that can be used to generate binnings satisfying all the three requirements of Section~2.

Given a region $\Omega$ and a set of points $\{z_{i}\}_{i=1}^{N}$, called \emph{generators}, in $\Omega$, a VT is a partition of $\Omega$ into regions $\{V_{i}\}_{i=1}^{N}$ enclosing the points closer to $z_{i}$ than to any other generator. Each $V_{i}$ is referred to as the \emph{Voronoi region} or \emph{bin} associated to $z_{i}$ (see Okabe, Boots, \& Sugihara 1992, for a comprehensive treatment).

The VT presents many interesting features for the binning problem:
\begin{itemize}
\setlength{\itemsep}{0cm}
\item it naturally enforces the Topological requirement;

\item it is efficiently described by the coordinates of its generators;

\item it is very easy to implement in the discrete case: given the generator positions, it is sufficient to locate the closest generator to any given pixel to determine the bin to which it belongs.
\end{itemize}

On the other hand, the fact that a VT is adopted for binning does not
enforce by itself the Morphological requirement: the bins are convex
by construction, but can have very sharp angles. Furthermore, the
Uniformity requirement is not addressed in any way by the use of a VT.
These two requirements have to be tackled through a properly tailored
distribution of the Voronoi generators. We present now a way to
produce such a distribution.

\subsection{Centroidal Voronoi Tessellation}
\label{sec:cvt}

The \emph{Centroidal Voronoi Tessellation} (CVT) is a technique which can be used to generate an optimally uniform \emph{and} regular VT in the continuous case, or in the limit of a large number of pixels. We initially assume that the observed signal $S_i$ can be described by a continuous function  $\rho({\bf r})$ in the sky-plane. Moreover we assume the noise $N_i$ to be a monotonic function of $S_i$ (e.g., Poissonian noise $N_i=\sqrt{S_i}$). With these assumptions the problem of generating equal-\SN\ bins reduces to that of producing a tessellation enclosing equal-mass, according to the density distribution $\rho({\bf r})$.

\begin{figure}[t]
    \plottwo{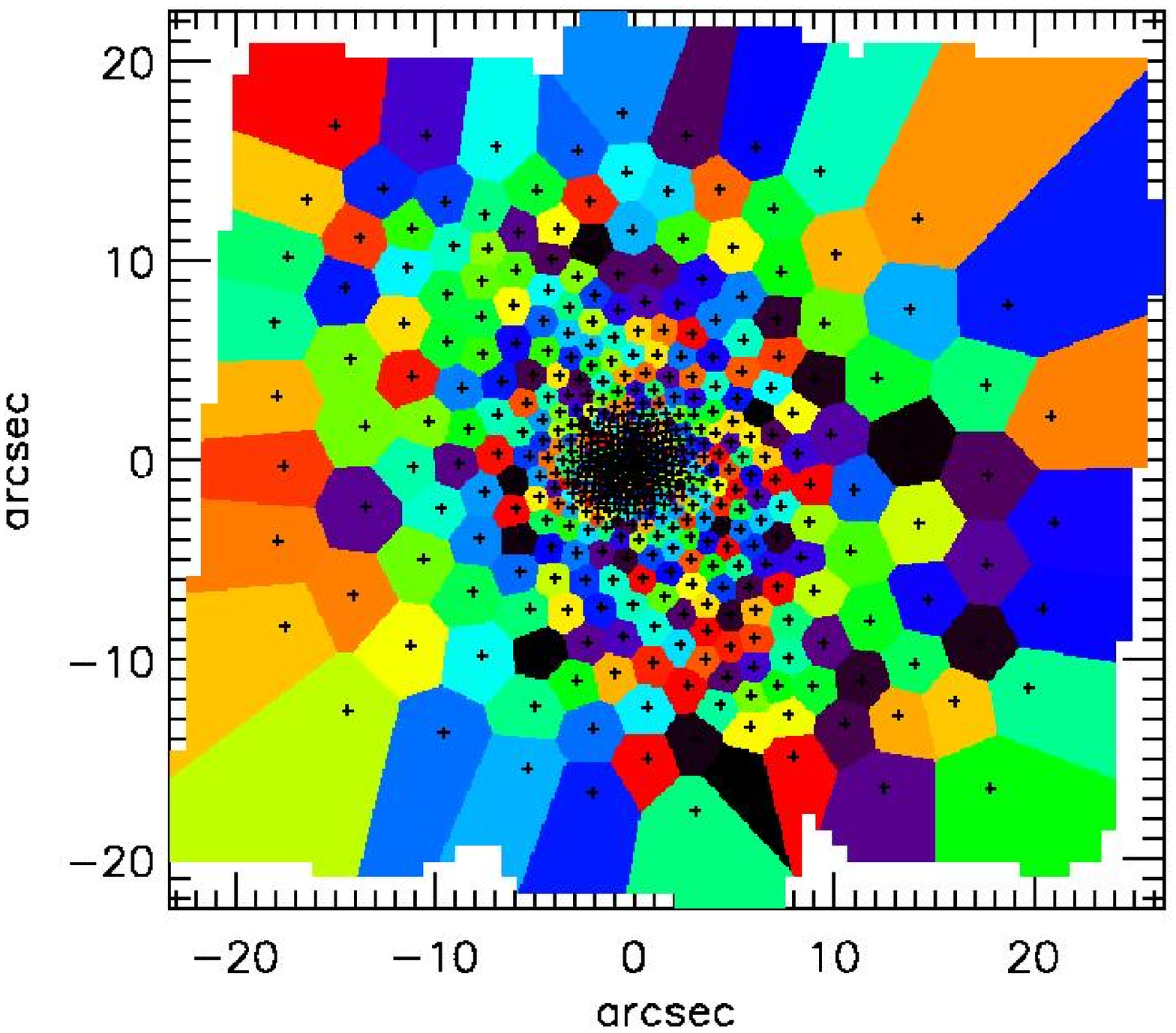}{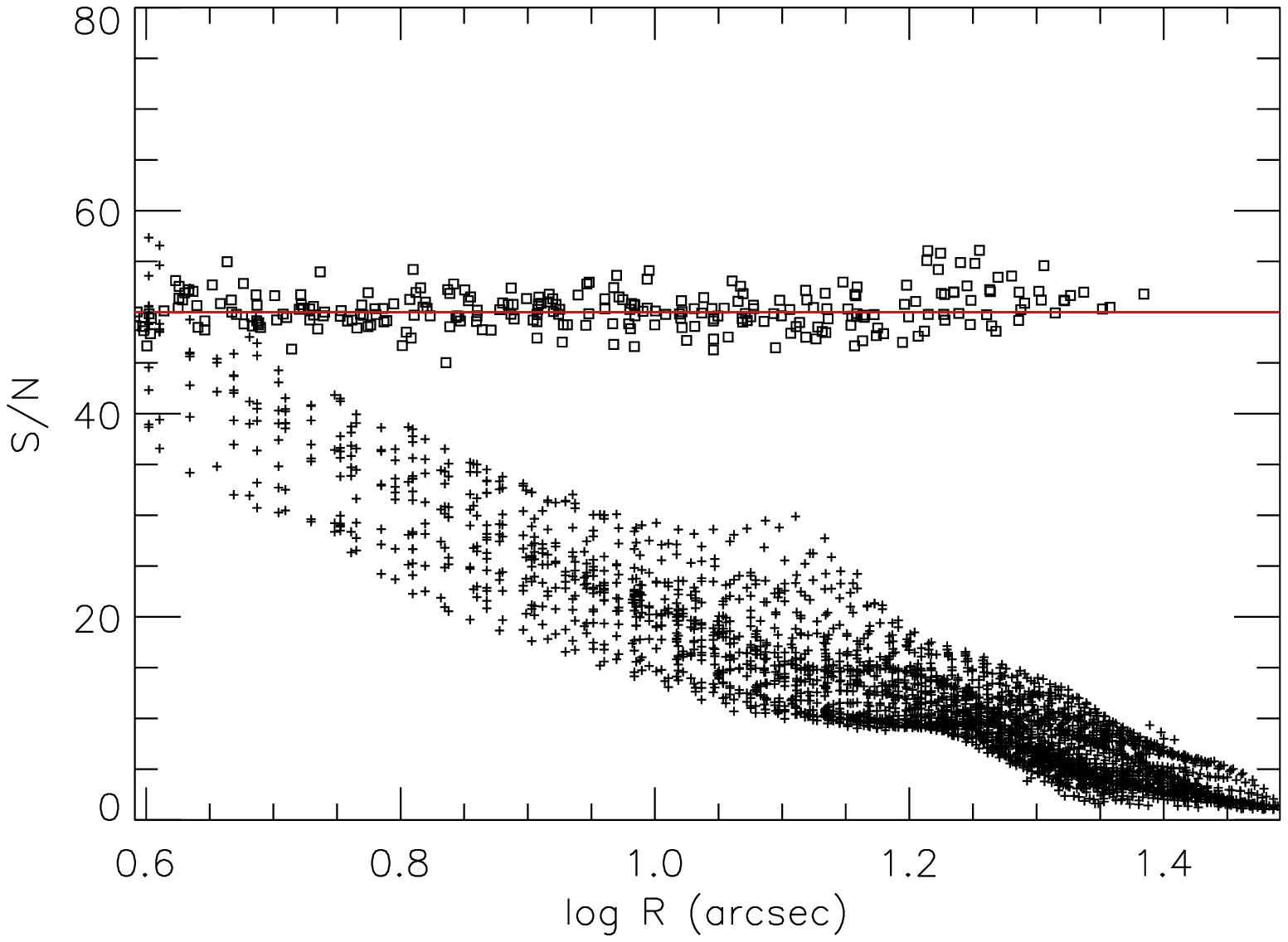}
    \caption{\emph{Left panel:} CVT-binning of the density $\rho^{2}({\bf r})$, where $\rho$ is obtained by linear interpolation from the surface brightness in \reffig{fig:n2273-SNmap}. This interpolation is used to `simulate' an observation with very high spatial resolution. The VT generators are shown with the crosses. \emph{Right panel:} \SN{} scatter in the CVT-binning of NGC~2273. The original \SN\ of the pixels is shown with the crosses, while the open squares represent the \SN\ of the final bins. The target \SN$=$50 is indicated by the red horizontal line.\label{fig:n2273-cvt}}
\end{figure}

Given a density function $\rho({\bf r})$ defined in a region $\Omega$, a CVT of $\Omega$ is a special class of VT where the generators $z_{i}$ happen to coincide with the mass centroids $z_{i}^{\ast} = \int_{V_{i}} {\bf r}\rho({\bf r})\,{\rm d}{\bf r}/\int_{V_{i}} \rho({\bf r})\,{\rm d}{\bf r}$ of the corresponding Voronoi regions $V_{i}$. As illustrated in the review by Du, Faber, \& Gunzburger (1999), the CVTs are useful to solve a variety of mathematical problems, but can also be observed in many real-life natural examples (living cells, territories of animals, etc.).

One of the most striking characteristics of CVT in the 2D case is its ability to partition a region into bins whose size varies as a function of the underlying density distribution, but whose shape tends asymptotically to a hexagonal-like lattice for a large number of bins. Another nice feature of CVT is that a simple algorithm exists for its practical computation: the Lloyd (1982) method, for which the CVT is a fixed point.

Although CVT bins are naturally smaller where the density is higher, the area-$\rho$ relation of the bins is not such that the mass enclosed in every bin is the same: the CVT cannot be used directly to produce equal mass bins (equal \SN{} in the case of photon noise). However, it can be shown that if a CVT is constructed for the density $\rho^2$, the tessellation obtained will enclose equal mass according to the original density $\rho$ (see Cappellari \& Copin in prep.\ for details).

\reffig{fig:n2273-cvt} presents a CVT produced by applying the Lloyd's algorithm to the density $\rho^2$, where $\rho$ was obtained by linearly interpolating the surface brightness of \reffig{fig:n2273-SNmap} onto a grid with pixel size $8\times$ smaller than the original size. This interpolation is used here to `simulate' an observation with a much higher spatial resolution, but is not an accurate way of dealing with lower resolution data. In this case of a large number of pixels the cells of the VT tend to the theoretical hexagonal shape and adapt nicely to density variations and to the irregular boundary of the region. The scatter of the \SN{} is also close to optimal with RMS scatter of $\sim 4\%$.

\begin{figure}[t]
    \plottwo{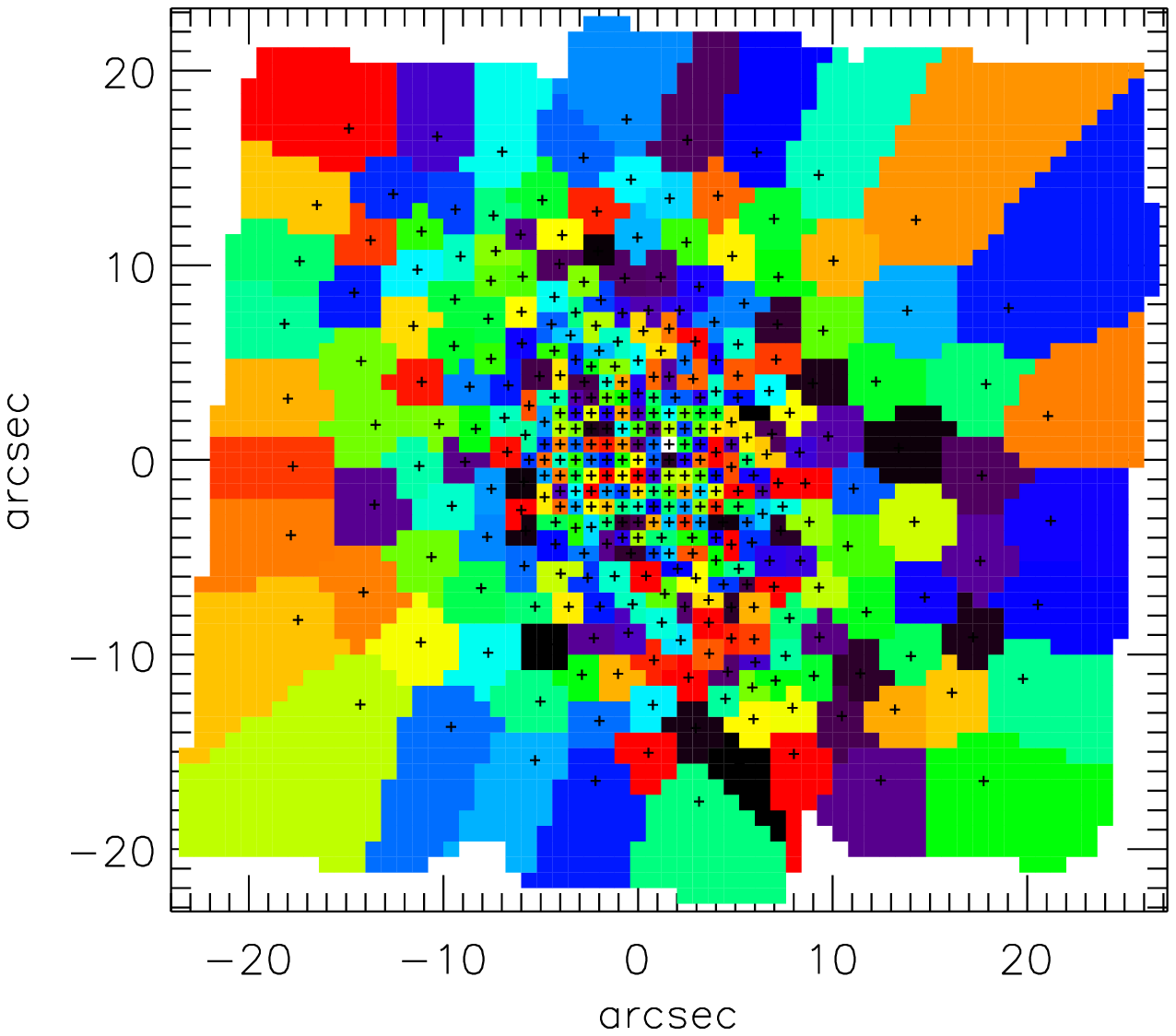}{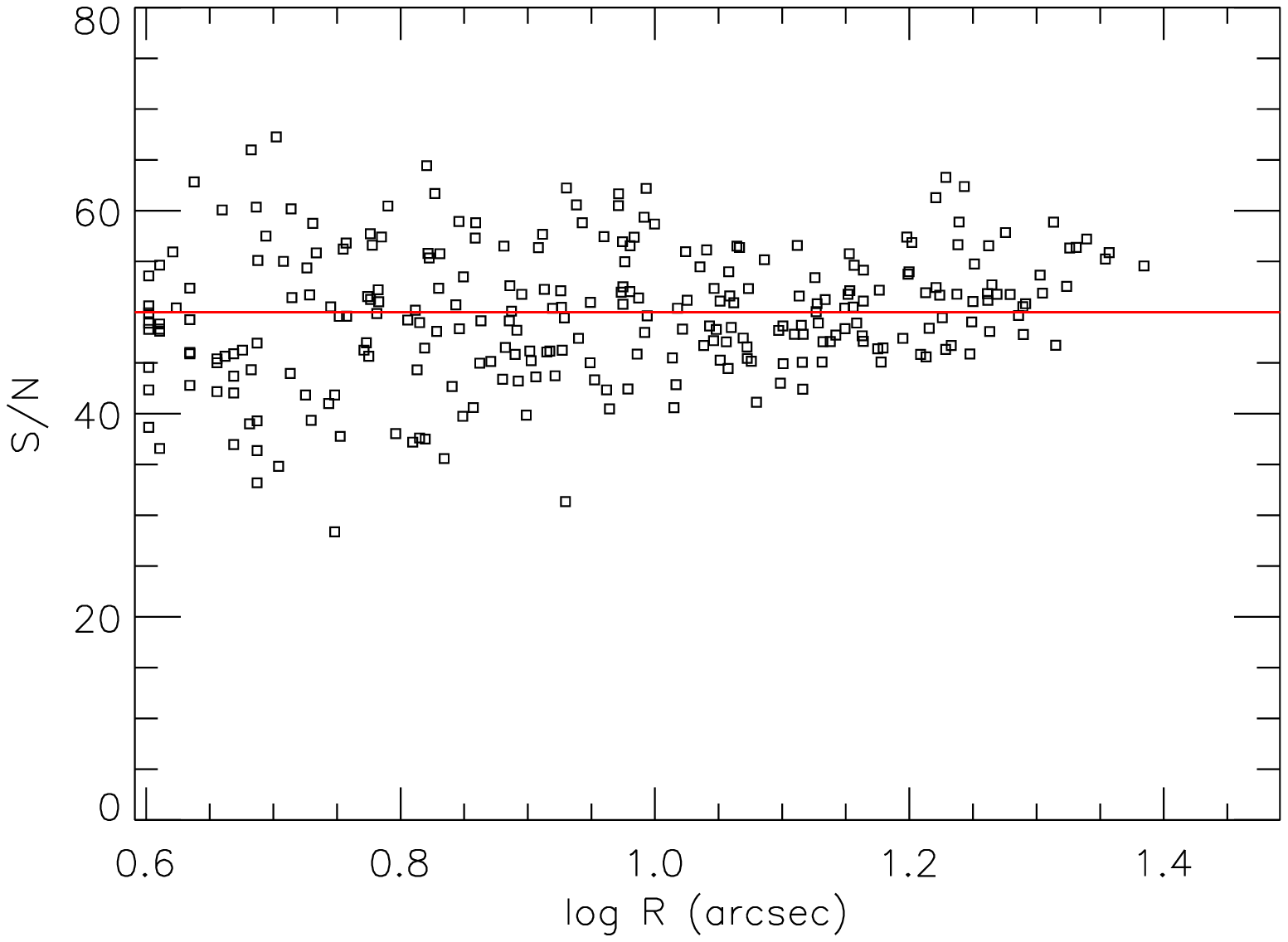}
    \caption{\emph{Left panel:} VT of the \sauron\ pixel grid of \reffig{fig:n2273-SNmap}, obtained using the same generators as in \reffig{fig:n2273-cvt}. \emph{Right panel:} \SN{} scatter in the bins as a function of the distance of the bin centroid from the galaxy center. Note the significant increase of the scatter, compared to \reffig{fig:n2273-cvt}. \label{fig:n2273-cvt-bad}}
\end{figure}

This CVT method illustrates the goals towards which an optimal  2D-binning algorithm should tend, but it has still some practical limitations:
\begin{itemize}
\setlength{\itemsep}{0cm}

\item it generates equal-mass bins and \emph{not} necessary equal-\SN{} bins, unless the noise is a monotonic function of the signal. It is e.g., useful to produce equal \SN{} bins when all the noise in the pixel can be attributed to photon noise. This is often the case, but not always;

\item more importantly the method does not work well when the bins are constituted of just a few pixels. In \reffig{fig:n2273-cvt-bad} the same generators as in \reffig{fig:n2273-cvt} where used to construct a VT for the coarser \sauron\ pixel grid of \reffig{fig:n2273-SNmap}. The obtained VT is similar to that of \reffig{fig:n2273-cvt}, but the \SN\ scatter increases considerably, due to discretization effects.
\end{itemize}

In practice an optimal 2D-binning method has to preserve the good characteristics of the CVT, in the limit of many pixels, but has to be able to take the discrete nature of pixels into account, when dealing with bins constituted by just a few pixels. This algorithm is the subject of the next section.

\subsection{Bin-Accretion Algorithm}
\label{sec:accretion}

We describe a method we developed to find the generators for the optimal VT of an image or IFS observation taking the discrete nature of pixels into account from the beginning. The algorithm described here constructs an initial binning trying to generalize to 2D the standard pixel-by-pixel binning algorithm used in 1D. The centroids of the bins found in this way are then used as starting generators for a CVT. The method reduces to the previous CVT in the limit of many pixels, but works on a pixel basis with bins made by just a few pixels.

A natural 1D-binning algorithm starts a bin from the highest \SN{} unbinned pixel and accretes pixels until a given target \SN{} is reached. To extend this idea to 2D, we need to make a choice for the direction toward which new pixels are accreted to a 2D-bin. The adopted method always tries to add to the current bin the pixel that is closest to the bin centroid. Furthermore, when a new bin is started, the first pixel is also selected as the one closest to the centroid of all the previously binned pixels. This simple scheme automatically \emph{tends to} generate bins that are compact, and the bin \SN{} can be carefully monitored on a pixel-by-pixel basis during the accretion phase. Some small imperfections remain at the end of the accretion phase: these are corrected by performing a CVT, using as starting generators the centroids of the previous bins.

\begin{figure}[t]
    \plottwo{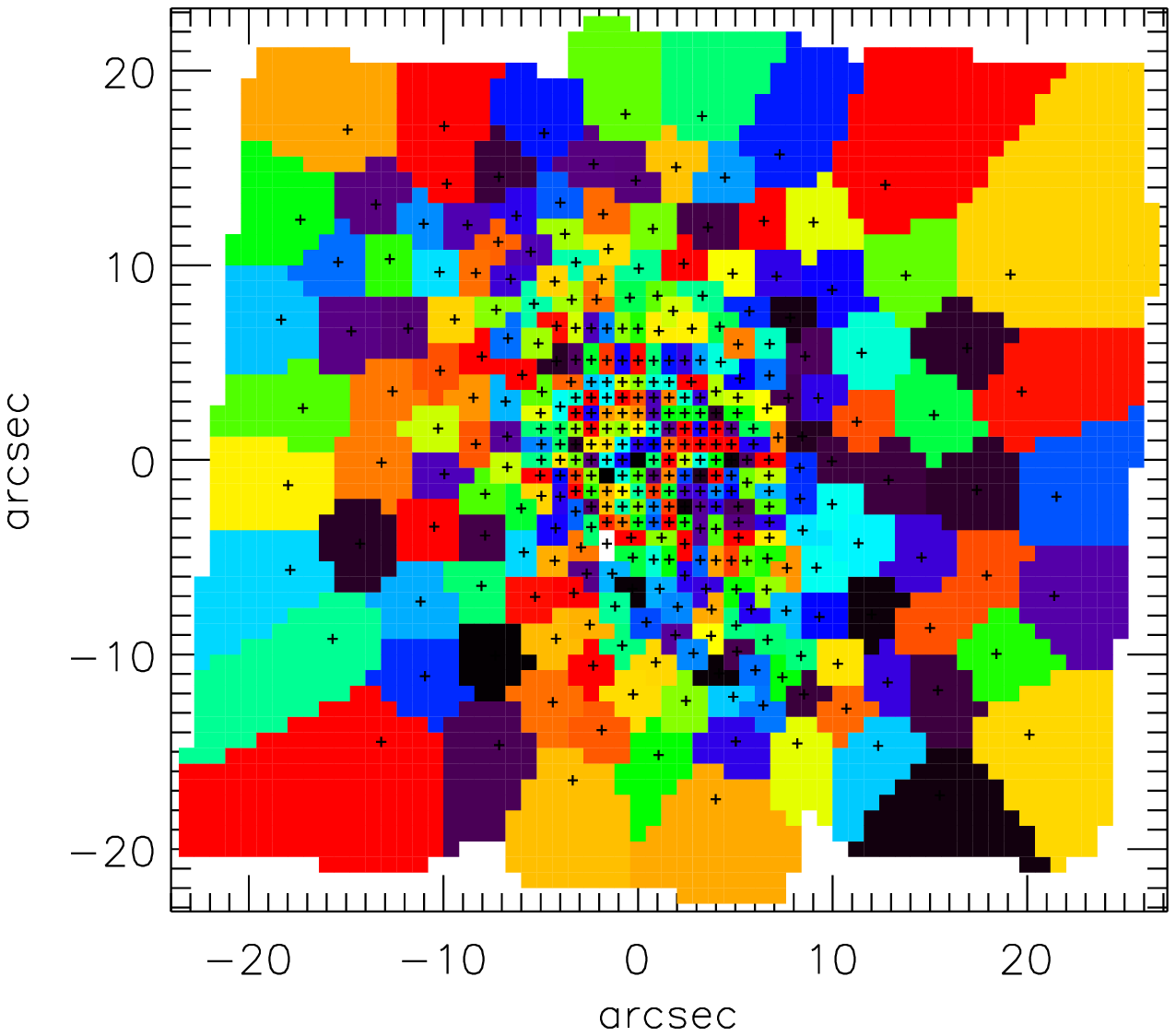}{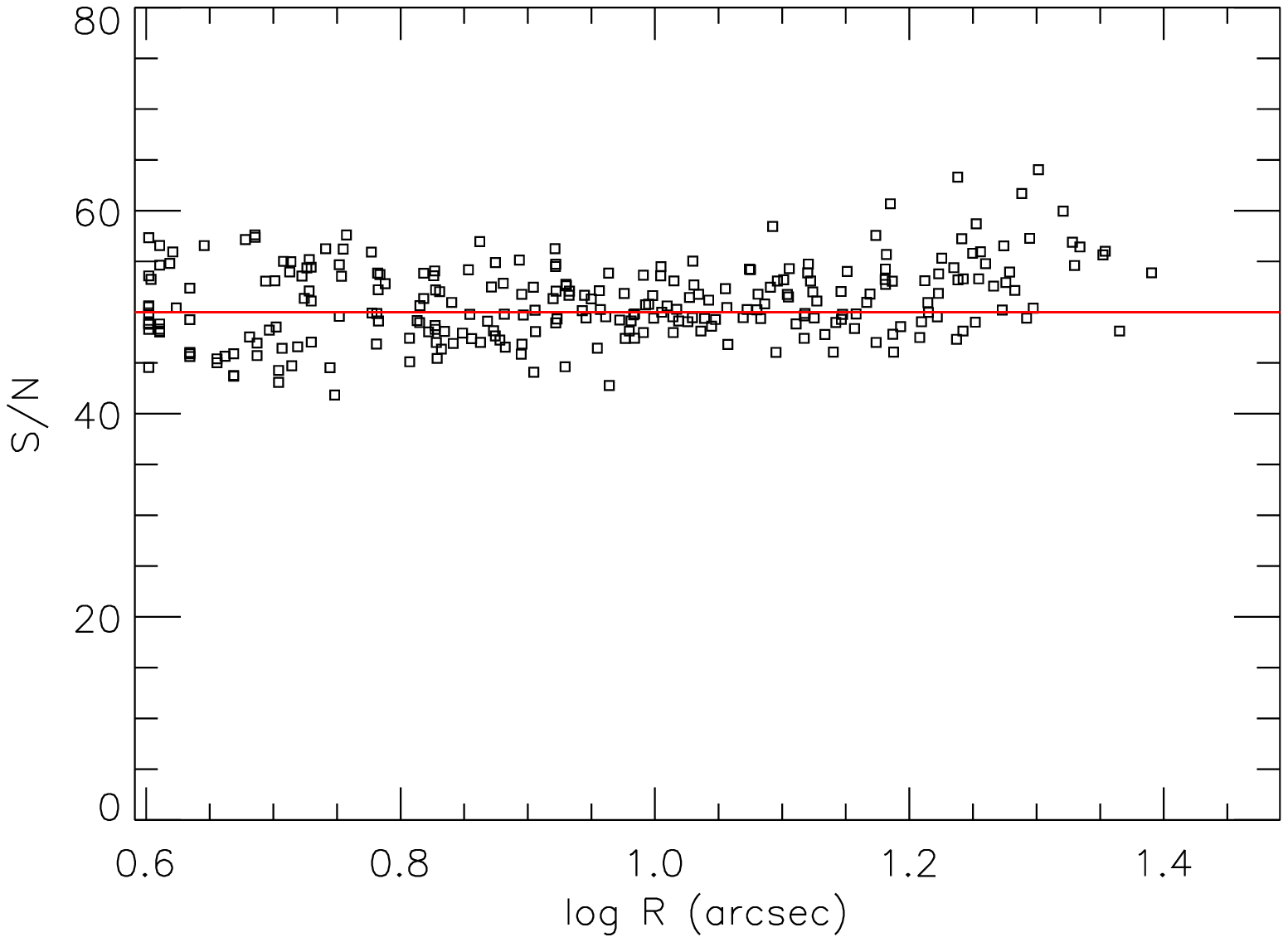}
    \caption{\emph{Left panel:} Final result for the bins after application of the bin-accretion algorithm to the \sauron\ \SN{} map of NGC~2273. \emph{Right panel:} the bins \SN\ as a function of the distance from the galaxy center. Note the decrease of the \SN\ scatter compared to \reffig{fig:n2273-cvt-bad}. \label{fig:bubbles-final}}
\end{figure}

\begin{figure}[t]
    \plottwo{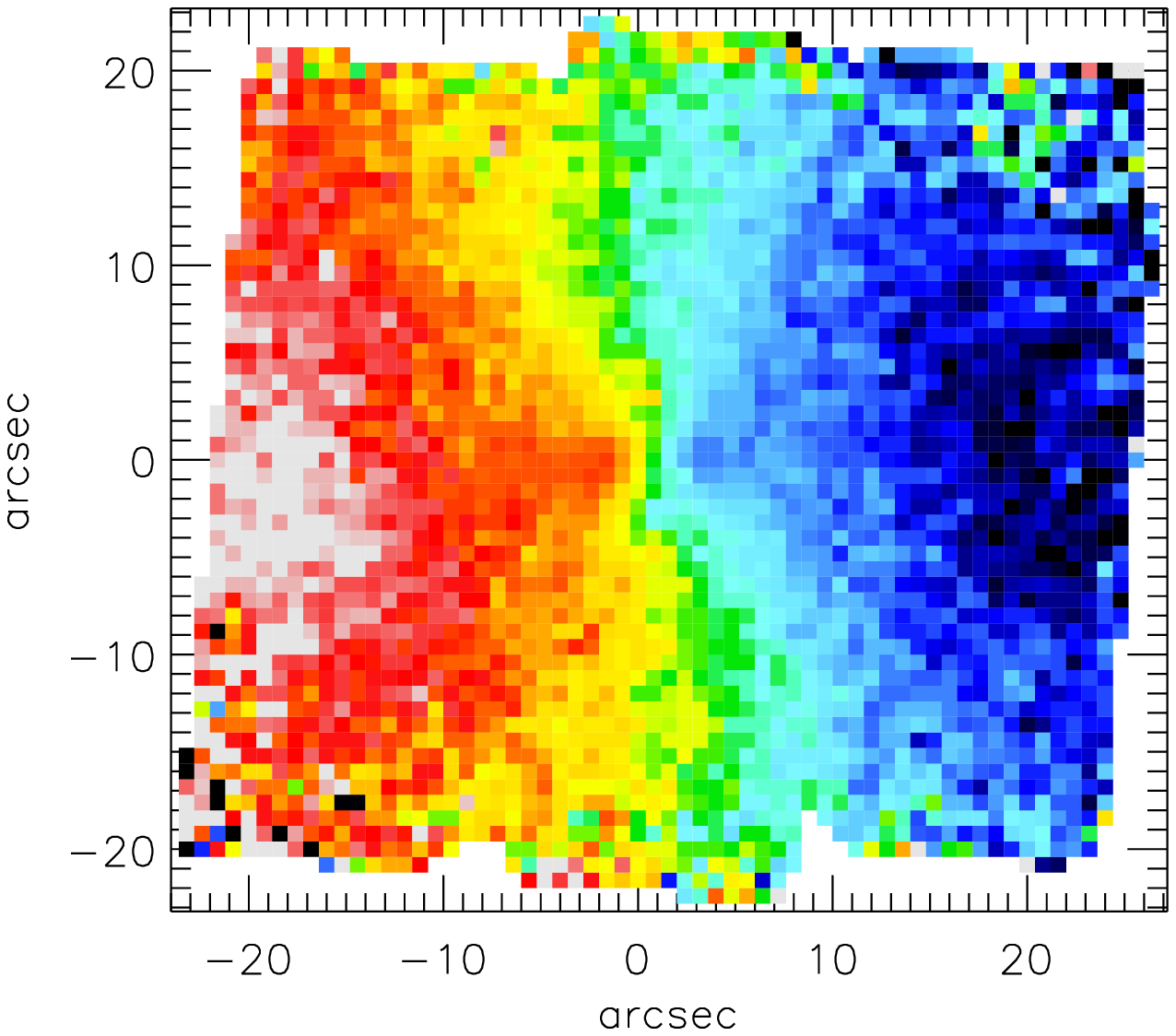}{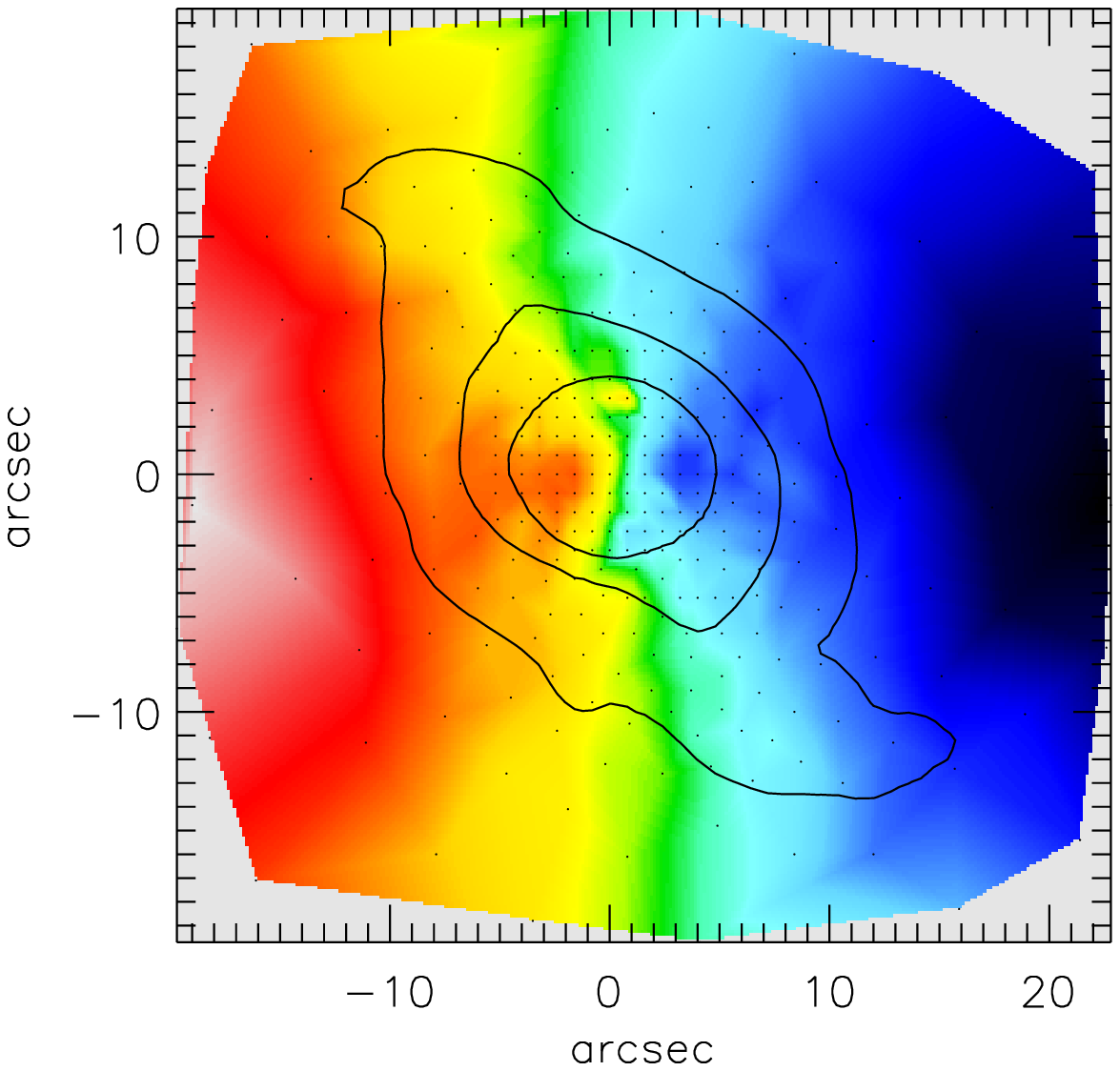}
    \caption{The stellar mean velocity field measured from the unbinned \sauron\ data of NGC~2273 (\emph{left panel}), is compared to the interpolated velocities extracted from the 2D-binned spectra (\emph{right panel}). The centroids of the bins used (\reffig{fig:bubbles-final}) are shown with the black dots. Some representative contours of the galaxy surface brightness are also shown. \label{fig:n2273_velfield}}
\end{figure}

An example of the application of the bin-accretion algorithm to the binning of the actual \sauron\ data of NGC~2273 is shown in the left panel of \reffig{fig:bubbles-final}, while the resulting \SN{} scatter is shown in the right panel: the RMS value is $\sim 6\%$. The \SN{} values, symmetrically clustered around the target $\SN{} = 50$, essentially represent the lowest \SN{} scatter obtainable from a binning of these data: all the scatter is due to discretization noise, which increases towards to the galaxy center, where the bin are made of a smaller number of pixels. The velocity field extracted from the binned data is shown in \reffig{fig:n2273_velfield}.

The binning resulting from this algorithm is qualitatively similar to the one obtained using the CVT (see \reffig{fig:n2273-cvt}). But, by contrast to the CVT alone, this method is able to produce bins that are essentially optimal also in the `small-bins' regime, with bins of only 2--4~pixels.

\vspace{-0.2cm}
\section{Conclusions}
\vspace{-0.2cm}
\label{sec:concl}

We presented a method to adaptively perform spatial binning of 2D data (e.g., IFS or imaging data). As an example of its application we have extracted the stellar mean velocity field from the 2D-binned \sauron\ data of the barred spiral galaxy NGC~2273. Adaptive 2D-binning should become common practice for the analysis of 2D data (in particular spectral data), as it is for 1D observations.

\acknowledgements

We thank Tim de Zeeuw for useful comments on the draft and Gijs Verdoes Kleijn for lively discussions on this subject.


\vspace{-0.2cm}
\begin{references}
\vspace{-0.2cm}

\reference Bacon, R., et al.\  2001, \mnras, 326, 23

\reference Du, Q.,  Faber, V., \& Gunzburger, M.\  1999, SIAM Review, 41, 637

\reference Kuijken, K., \& Merrifield, M.~R.\ 1993, \mnras, 264, 712

\reference Lloyd, S.~P.\ 1982, IEEE Transactions on Information Theory, 28, 129

\reference Okabe, A.,  Boots, B., \&  Sugihara, K.\ 1992, Spatial tessellations concepts and applications of Voronoi diagrams (Chichester, UK: Wiley)

\reference Rix, H.-W., \& White, S.~D.~M.\ 1992, \mnras, 254, 389

\reference Samet, H.\ 1984, The Quadtree and Related Hierarchical Data Structures. ACM Computing Surveys, 16, 187

\reference Sanders, J.~S.,  \& Fabian, A.~C.\  2001, \mnras, 325, 178

\reference van der Marel, R.~P., \&  Franx, M.\  1993, \apj, 407, 525

\end{references}
\end{document}